# Does Oversizing Improve Prosumer Profitability in a Flexibility Market? - A Sensitivity Analysis using PV-battery System

Babu Kumaran Nalini, Michel Zade, Zhengjie You, Peter Tzscheutschler and Ulrich Wagner

Chair of Energy Economy and Application Technology,
Technical University of Munich, 80333 Munich, Germany

**ABSTRACT**

The possibilities to involve small-scale prosumers for ancillary services in the distribution grid through aggregators and local flexibility markets, question whether it is profitable for prosumers to oversize proactively. In this analysis, a Python model is developed to identify the cost optimal operation plan of the PV-battery system and to evaluate the device flexibility. An economic assessment is carried out to derive the cost-benefit of sizing based on the mean electricity price. A sensitivity analysis is performed with the above results to study the profitable sizing of PV-battery systems with flexibility services. The results show promising advantage of oversizing although limitations prevail with the extent of flexibility services offered.

## 1. INTRODUCTION

The increasing integration of distributed energy resources (DERs) in the distribution grid provokes measures to coordinate energy sufficiency and control uncertainty. Recent studies have proposed a variety of techniques to maintain real time load and generation coordination between distribution system operators (DSOs), utilities and prosumers. Concepts such as energy aggregators, local energy markets (LEMs), peer-to-peer energy trading have been the topics of interest to allow energy sufficiency with the communities [1–4].

Fortunately, it is possible for the DSOs to take advantage from the prosumer and procure flexibility services to serve ancillary services and congestion management [5]. Fonteijn et al. demonstrates a generic four step congestion management approach using prosumer flexibility [6]. Several methods have also been proposed to quantify the flexibility from prosumers devices such photovoltaics (PV) and battery system, electric vehicles, heat pumps to allow interaction of prosumers with the DSOs using flexibility markets [7–9].

The prosumer's participation to flexibility markets is motivated by the investment returns from flexibility services. In this paper, a case study is developed using PV-battery system to understand whether oversizing improves prosumers profitability when offering flexibility services. A sensitivity analysis is performed by estimating the flexibility of the prosumer devices and analyzing the mean price of electricity using an economic assessment.

The paper is structured such that the section 2 discusses the material and methods used to compute the optimal operation of the PV-battery system and flexibility estimation. Further, an economic assessment is carried out for the cost-benefit of sizing. In section 3, sensitivity analysis is detailed to identify the profitable sizing for the prosumers system with flexibility services.

## 2. METHODOLOGY

An operational plan of a residential PV-battery system is realized using a variety of historical and standardized datasets. Further, the optimal operation is characterized as a mixed integer linear programming (MILP) problem and is solved using a Python solver. A device specific flexibility estimation is performed for PV and battery energy storage system (BES). Based on [10], an economic assessment is carried out for the PV-battery system to determine the annual mean electricity price of the system configuration. Finally, a sensitivity analysis is performed to analyze the profitable PV-battery system configuration for prosumer participating in local flexibility markets (LFMs).

## 2.1 Input data

To evaluate the optimal operation of the PV-battery system, the time series data of the PV power generation ($p_t^{pv}$) is required. In this study, the historical measurement data during the year 2019 is used from the rooftop PV installation located at the Technical University of Munich, Germany (48.1507°,11.5695°). The total PV installed capacity of this configuration is 3kWp.

The electricity load data is obtained from the standardized household data published by the HTW Berlin university [11]. A randomly chosen single-family household is used for this study. The annual household electricity demand was found to be 8kWh.

The electricity import price $k_t^{imp}$ from the grid follow the day-ahead price trends of 2019 with a mean value of 0.31€/kWh including value added taxes [12]. The export price $k_t^{exp}$ to the grid was set as a fixed cost of 0.08€/kWh [13]. All the datasets used have a 15-minute time interval and are available for 365 days.

## 2.2 Cost-optimal operation

The cost-optimal operational schedule of the PV-battery system can be obtained by solving the prosumer device configuration as a MILP problem. The net cost $k$ transacted by the prosumer is a minimization problem over the day, with $\Delta T = 15$ min. Here, $p_t^{imp}$ and $p_t^{exp}$ corresponds to the net power imported or exported.

$$k = \min \sum_{t=1}^{T} (p_t^{imp} k_t^{imp} - p_t^{exp} k_t^{exp}) \Delta T \quad (1)$$

s.t.
$$p_t^{pv} + p_t^{bes,dh} + p_t^{imp} = p_t^{exp} + p_t^d + p_t^{bes,ch} \quad (2)$$
$$0 \leq p_t^{\phi} \leq \overline{p}^{\phi} \quad (3)$$
$$0 \leq e_t^{\phi} \leq \overline{e}^{\phi} \quad (4)$$
$$e_t^{es} = e_{t-1}^{es} + (p_{t-1}^{es,ch} \eta^{es,ch} - p_{t-1}^{es,dh}/\eta^{es,dh}) \Delta T \quad (5)$$

The constraints to solve this problem includes the electricity balance as shown in equation 2 where $p_t^{bat,dh}$ and $p_t^{bat,ch}$ correspond to the charging and discharging power of the battery system respectively, at any specific time step. Equation 3 and 4 is a simplified version of power and energy limit constraint which can be applied to both PV and battery charging/discharging $\phi \in [PV, BES]$. The energy stored $e_t^{es}$ in the battery depends on the state of charge from the previous time step and considers the efficiency of the charging $\eta^{es,ch}$ and discharging $\eta^{es,dh}$ process as shown in equation 5. A binary condition is used to any avoid simultaneous charging and discharging by the solver.

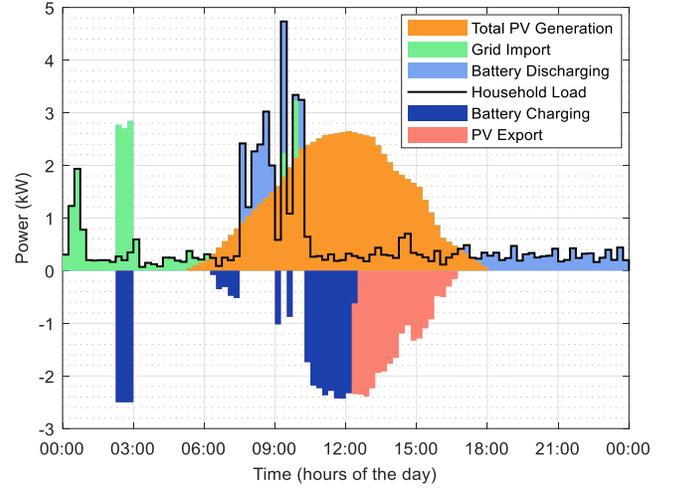

Figure 1 Optimal operation schedule with PV-BES systems [9]

Figure 1 shows the cost optimal operational schedule to carry out PV export and battery charging/discharging using which, the total import and export costs can be calculated. The optimization problem is solved using Python and GLPK solver. The simulation takes about 30 seconds to solve 96 steps using an Intel I7 processor.

## 2.3 Flexibility

Considering the optimized operation schedule as the baseline, the flexibility of the device is quantified by calculating all the possible deviation (throttling) without violating the device constraints. Since both upward and downward throttling is possible, the flexibility is defined into two types. Table 1 refers to the definitions of flexibility types used in this study.

Table 1 Flexibility definition

| | |
|---|---|
| Negative flexibility | Removal of energy from the grid |
| Positive flexibility | Addition of energy to the grid / Refrain a scheduled withdrawal |

Negative flexibility is used when excess energy is available in the grid and the prosumer's devices are requested to charge ahead of the schedule as flexibility service. Positive flexibility is used when the grid faces any deficit of energy and the prosumer feeds-in additional energy or refrains from withdrawing any scheduled grid charging. Typical example of these flexibility types is curtailment of PV feed-in as negative flexibility service $p_{t,flex-}^{pv}$ and the positive flexibility service $p_{t,flex+}^{bes}$ which corresponds to refrain from charging the battery at scheduled time step. A basic representation of the flexibility power calculation of the PV and battery systems are shown using the equations 6-8.

$$p_{t,flex^-}^{pv} = p_t^{pv} \quad (6)$$

$$p_{t,flex^-}^{bes} = \overline{p}^{bes,ch} + p_t^{bes,dh} - p_t^{bes,ch} \quad (7)$$

$$p_{t,flex^+}^{bes} = \overline{p}^{bes,dh} - p_t^{bes,dh} + p_t^{bes,ch} \quad (8)$$

The duration of the flexibility and the energy offered by the device can be calculated using equations 9 and 10 respectively. The duration of flexibility is the minimum time period in which a constant flexibility power can be offered. The flexibility energy is simply calculated as the product of power and duration along with the step size.

$$\delta_{t,flex}^{\phi} = N\left(p_{n,flex}^{\phi} \geq p_{t,flex}^{\phi}\right) \quad \forall n \in [t+1, T] \quad (9)$$

$$e_{t,flex}^{\phi} = p_{t,flex}^{\phi} \cdot \delta_{t,flex}^{\phi} \cdot \Delta T \quad (10)$$

The flexibility equations are solved considering the device constraints such as SOC limits, charging limits, etc. A detailed understanding on the flexibility quantification can be gathered from [9].

*2.4 Economic assessment*

The economic assessment of sizing the PV-battery system involves closer understanding of the energetic assessment of the setup. Weniger et al. has proposed an assessment method to calculate the economic and energetic properties, required for analyzing the size of the PV-battery system [10]. The capital expenditures of the devices are represented as annual cost $C^{\phi}$ with $\phi \in [PV, BES]$ and it is calculated using equation 11. Here, $\tau^{\phi}$ corresponds to the investment cost of the device which is multiplied by the capacity or system size $\overline{p}^{\phi}$.

$$C^{\phi} = \overline{p}^{\phi} \cdot \tau^{\phi} \cdot (\alpha_{\phi} + \beta_{\phi}) \quad (11)$$

$$\alpha_{\phi} = \frac{r}{1-(1+r)^n} \quad (12)$$

Annuity factor $\alpha_{\phi}$ is used to describe the annual payments with rate of interest (r) of the investment for a defined period of investment n as shown in equation 10. The operating expenditure $\beta_{\phi}$ in this study is considered as 4% of the capital expenditure. Two important ratios are required to calculate the annual expenditure and revenue of the prosumer's energy transaction with the grid, namely export ratio $f_e$ and self-sufficiency $f_s$.

$$f_e = \frac{E^{pv,exp}}{E^{pv}} \quad (13)$$

$$f_s = \frac{E^{d,pv} + E^{d,bes}}{E^d} \quad (14)$$

Export ratio is the amount of annual feed-in energy $E^{pv,exp}$ from the PV system to the grid relative to the PV annual energy yield $E^{pv}$. The self-sufficiency ratio is the share of prosumers demand which is satisfied by the PV system $E^{d,pv}$ as well as from the battery discharge process $E^{d,bes}$ relative to the annual energy demand $E^d$ of the prosumer. The annual expenditure of the prosumer for electricity import from the grid is calculated using equation 15.

$$C^{el} = E^d \cdot k_t^{imp} \cdot (1 - f_s) \quad (15)$$

The annual revenue for the prosumer is characterized into two types, a direct revenue from PV feed-in and probable revenue from flexibility services. The feed-in revenue is computed as a product between the feed-in price and annual PV feed-in energy as per 16.

$$R^{pv} = E^{pv} \cdot k_t^{exp} \cdot f_e \quad (16)$$

The revenue from the flexibility services is not fixed as the flexibility services are called only upon on request after submitting bids to the LFMs. The annual average flexibility profit and number of flexibility services offered are $k_t^{flex}$ and $\vartheta$ respectively. The flexibility energy $E^{\phi,flex}$ is assumed to be an average of positive and negative flexibility since only one of the two flexibility types can be offered at any instance.

$$R^{flex} = E^{\phi,flex} \cdot k^{\phi,flex} \cdot \vartheta \quad (17)$$

Using the investment costs, expenditures and revenue, it is possible to calculate the mean price of electricity transacted $\hat{C}$ by the prosumer. Equation 18 is used to obtain the mean price in €/kWh by calculating the total cost relative to the total demand.

$$\hat{C} = \frac{\sum C^{\phi} + C^{el} - R^{pv} - R^{flex}}{E^d} \quad (18)$$

The annual mean of electricity price (AMEP) signifies the profitability of a specific sizing of the PV-battery system. By varying the size of the PV-battery system and the amount of flexibility services offered annually, it is possible to determine a specific device configuration which is profitable for the prosumer to invest considering LFM interactions.

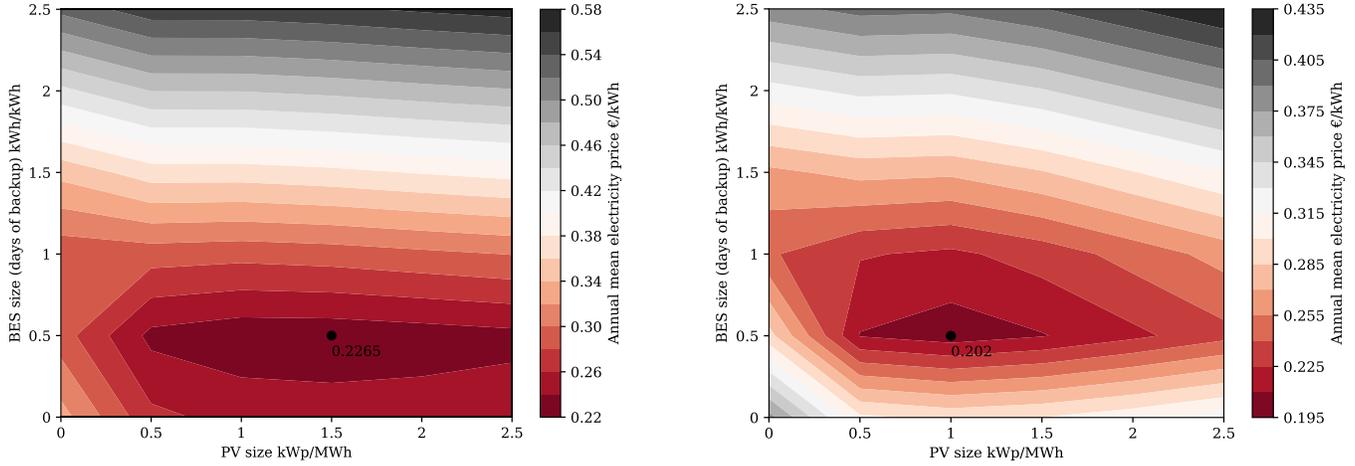

*Figure 2 Annual mean electricity price for PV-BES configurations considering present (left) and long-term (right) price developments*

## 3. SENSITIVITY ANALYSIS

To understand the impact of PV-battery system sizing on the mean price of electricity, a sensitivity analysis is performed using two case studies considering the price developments in the present and long-term scenario.

### 3.1 Assumptions

The rate of interest for the annuity factor is assumed to be 3.5% while considering the lifetime of the PV and BES systems to be 20 years. The investment cost of the systems used in both the case studies are shown in Table 2 and the assumptions are based on [14] considering long-term values to be year 2040. The import electricity price is set to increase at a rate of 1.5% annually and is assumed to have a mean of 0.37€/kWh in the long-term. The feed-in electricity price has been decreasing annually and is assumed to reach 0.02€/kWh in the long-term.

*Table 2 Case-study parameters of PV and BES considering the system costs and price developments in the present and long-term scenario*

| Parameters | Present | Long-term |
|---|---|---|
| PV system cost (€/kWp) | 1200 | 800 |
| BES system cost (€/kWh) | 900 | 500 |
| Electricity import (€/kWh) | 0.319 | 0.37 |
| Feed-in tariff (€/kWh) | 0.068 | 0.02 |

Figure 2 shows the least cost-optimal sizing of the PV and BES in both the scenarios considering no flexibility revenues ($R^{flex} = 0$). The figure axes are shown as the fraction of size of PV (kWp) and annual electricity demand (MWh) for PV systems and as fraction of capacity of BES (kWh) with respect to the daily mean electricity demand (kWh) for the BES systems. Similar to the results shown in [10], it can been observed the optimal size for the present day scenario converges at AMEP = 0.226€/kWh. This implies profitable installation when larger PV system and smaller BES are used due to higher revenue from feed-in tariff. In contrary, in the long-term when the import prices increase and the feed-in tariff reduces the convergence happens at reduced PV size. It is to be understood that the convergence point only denotes the least cost-optimal configuration.

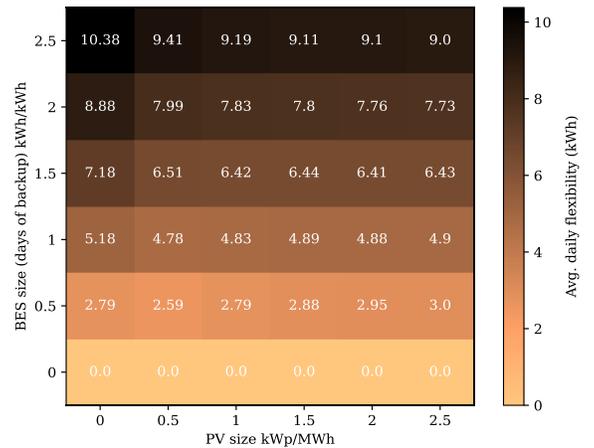

*Figure 3 Average flexibility energy for PV-BES configurations*

It is always possible for the prosumer to install higher BES systems for better reliability and still be profitable until the corresponding AMEP does not exceed the import electricity price. The average flexibility energy of a flexibility bid for different PV-battery system sizes are shown in Figure 3. It can be observed that increasing the size of PV-battery system increases the amount of flexibility energy offered for a flexibility service. When the BES charges from the PV system, the flexibility offered in these steps are reduced. When no PV system is involved, the BES system is highly flexible to charge from any time step since it has no choice but to only import from the grid. Whereas, when the PV system is available the BES tries to charge from PV to the maximum extent. Therefore, the average flexibility energy to grid slightly drops when the size of PV increase.

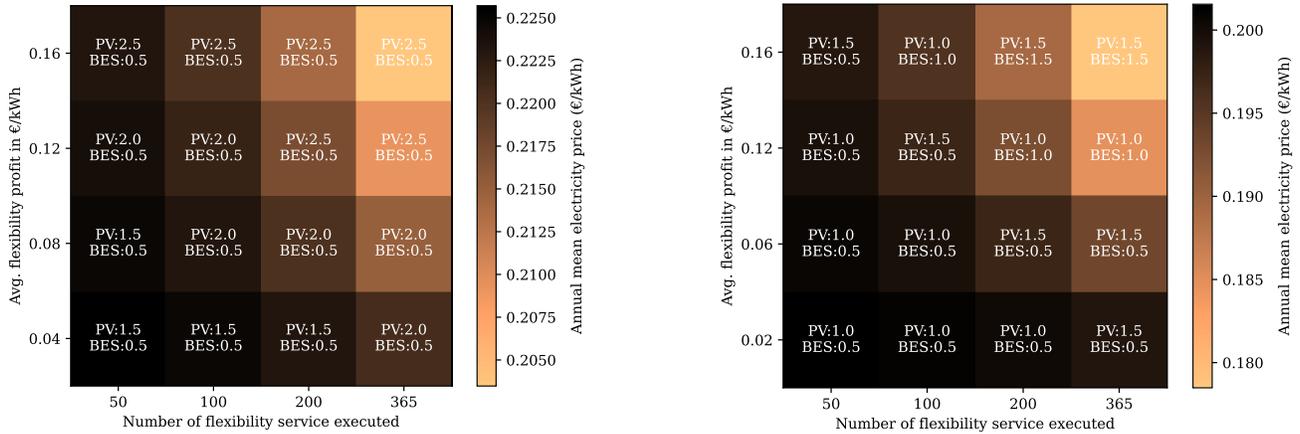

*Figure 4 Sensitivity of mean electricity prices to flexibility service considering present (left) and long-term (right) price developments*

Based on the figure 2 and 3, a sensitivity analysis is performed to find the impact on AMEP by varying the number of flexibility service offers executed and the average flexibility profit. The amount of flexibility served is equal to the average flexibility energy. Figure 4 shows the sensitivity of mean electricity prices with respect to the flexibility service in the present and long-term scenario. The sensitivity is obtained by deriving the least cost-optimal configuration as shown in figure 2 while varying the flexibility profit and the number of flexibility offer executed in a year. The size of the system satisfying the AMEP is annotated within the heatmap.

As the flexibility price and number of flexibility offers executed are unknown and depends on the grid situation and market execution, in this study, an analysis is performed over the possible range of values. It is assumed that the number of flexibility offers executed by the prosumer rarely exceeds more than one offer a day since the results from OpenTUMFlex show that the flexibility energy left after executing one flexibility service in a day reduces drastically for a small-scale prosumer. Therefore, the number of flexibility offers are limited to 365. The profit from flexibility is restricted between the levelized cost of electricity generation (LCOE) of the system to 0.16€/kWh. This maximum limit is an assumed considering the LCOE of gas turbines, combined cycle plants as per [14] since it would become easier to install exclusive devices to provide flexibility rather than aggregating prosumer flexibility potential. Nevertheless, there might exist cases where the full-load hour requirements are very less and it is not suitable to install exclusive devices. Here, the annual revenue from prosumer flexibility would also be lower even when the flexibility profit is higher than this study's assumed limit.

## 4. RESULTS

The impact of sizing on the prosumer profitability is analyzed through a sensitivity analysis on the annual mean electricity price with respect to a range of possible flexibility service offered by the prosumer. From figure 4, it can be deduced that over-sizing impacts prosumer profitability only when either the number of flexibility services and/or the profit from flexibility service are relatively high. The prosumer is able to profit sufficiently by optimally sizing the system with respect to figure 2.

The relative increase in the profit from flexibility happens only when the prosumer is able to execute significant amount of flexibility service throughout the year since the revenue from flexibility is relatively low in comparison to investment cost required for oversizing the systems. In the present scenario, the optimal sizing with flexibility is never achieved while increasing the BES size due to increased capital investment. Therefore, the over-sizing is dominated by the PV systems. In the long-term scenario, the BES system size plays a role with different levels of flexibility service due to increased import price and reduced investment and feed-in tariff.

This analysis performs a generic investigation to show the interdependencies between sizing, flexibility and profitability. The future research can focus on including several aspects such as to consider size specific capital investment costs or independently define different revenue for positive and negative flexibility. The disadvantage of this method is that when the prosumer demand changes over the year due to inclusion of new systems such as electric vehicle then the cost-optimal configuration will not remain the same. Also, this study is focused on PV-battery system and the flexibility services offered by other devices such electric vehicles or heat pumps may have different outcomes.

**ACKNOWLEDGEMENT**

The Federal Ministry for Economic Affairs and Energy (BMWi), Germany supports this contribution as a part of the SINTEG project C/sells under the grant number 03SIN109.

**SUPPLEMENTARY MATERIAL**

This analysis uses an in-house developed Python model called OpenTUMFlex for flexibility estimation. https://github.com/tum-ewk/OpenTUMFlex.